\documentstyle[12pt,preprint]{aastex}
\begin{document}

\title{The Emergence of Consciousness in the Quantum Universe} 

\author{Xiaolei Zhang}
\affil{Department of Physics and Astronomy\\
George Mason University\\
Fairfax, VA 22030, USA\\
xzhang5@gmu.edu}

\section*{ABSTRACT}

It is argued that human consciousness is likely to have emerged
during the self-consistent evolution of the physical universe,
through the gradual accumulation of biological entities' ability
to tap into the intrinsic non-deterministic potentiality in
the global nonequilibrium phase transitions occurring continually 
in the quantum universe.  Due to the fact that the matter and 
energy content participating in these global phase transitions is a continuum, 
there are in effect infinite degrees-of-freedom in the substratum, which 
invalidate the usual deterministic laws of the mechanical evolution, 
and allow chance factors to appear in the emergent properties of 
resonantly-formed quantum particles, especially in the acquired phases 
of their wavefunctions. Such chance factors, though occurring mostly 
randomly during the early cosmic evolution phase, can be harnessed more 
``purposefully'' by the biological entities co-evolving with the physical 
universe, due in part to the globally-entangled nature of quantum 
interactions. Over time, this ability to deliberately manipulate 
the chance factor in the fundamental level of quantum interactions 
gradually evolves into the free will and self-awareness of higher 
biological beings. The emergence of higher-level consciousness in turn 
greatly enhances the entropy-production ability of the biological entities,
 making them a powerful new form of the ``dissipative structures'' that 
nature constructs to accelerate the entropy evolution of the universe.
It is further shown that complex, nonequilibrium systems (including the 
universe as a whole) have the innate ability to spontaneously generate 
ever-renewable, highly-complex structural features in response to 
a changing and evolving environment; therefore natural selection is not 
the only driving mechanism for the irreversible evolution of the internal 
characteristics of biological species.

{\bf Keywoods:} Consciousness; quantum measurement; quantum entanglement; 
Mach's Principle; dissipative structures; nonequilibrium; evolution
phase transitions

\section{INTRODUCTION}

A number of researchers have previously suggested that quantum
processes might be playing an important role in the generation 
of human consciousness (see, e.g., Penrose et al. 1993 and the 
references therein; Stapp 2009 and the references therein).  
A main reason for investigating the role of quantum mechanics in 
consciousness formation lies in the realization that classical 
physics allows essentially only deterministic evolution -- meaning
that the details of all future events are predetermined from the 
initial conditions at the formation of the universe more than 10
billion years ago. Deterministic dynamics obviously could
not serve as the basis for free will, which is an important
component of higher consciousness.  In quantum physics, on the
other hand, there is a chance factor which is revealed both
through the probabilistic interpretation of the quantum wavefunction, 
and through the probabilistic element in quantum-measurement
wavefunction collapse.

The current work continues this well-known tradition of 
modern thinking, but goes a significant step further in suggesting 
that the origin of consciousness is to be sought in the common origin
of all physical laws in a globally-connected quantum universe.  This 
new view on the origin of consciousness is based on a new conceptual 
view of the foundations of quantum processes (Zhang 2007),
in which the origin of all physical laws and the properties of
fundamental particles are sought in a form of generalized Mach's
principle.  It postulates that the matter and laws in the physical universe
are co-selected, and they co-evolve in a global ``resonant cavity'' consisting of
the entire matter and energy content of the universe.  Quantum measurements
in this theory are global nonequilibrium phase transitions happening
either spontaneously, or induced by artificially-enforced boundary
conditions in the universe resonant cavity.  Quantum jumps occur
because the universe environment becomes unstable 
to the formation of new quantum ``modes'' under the changing boundary
conditions, much like in classical systems such as microwave oscillators, which
form intrinsic modes of spatial-temporal oscillations when a certain boundary
condition is enforced.  What distinguishes these quantum global phase transitions
from classical ones is that the classical nonequilibrium phase transitions
can in general be isolated in a small, finite region, whereas the quantum global
phase transitions we are referring to happen in the entire universe as a whole, 
which explains why, for example, a given species of fundamental particle generated from 
high-energy reactions always has exactly the same intrinsic characteristics 
no matter whether they are generated in a reactor in CERN, or in Brookhaven.  
Another characteristic distinguishing quantum interactions from classical ones 
is that quantum wavefunctions are globally distributed, and are always entangled 
to varying degree with the wavefunctions of all the other matter in the universe, 
causing the propagation of their mutual influences to happen superluminally.
These characteristics of quantum interactions turn out to have important 
implications on how a quantum-mechanically-based consciousness model operates.
We will further show in this paper that, in fact, there is 
no clear distinction between a quantum measurement process and a unitary 
evolution process. Quantum processes are unified by a continuous 
variation of characteristics bridging these two theoretically-idealized extremes.
This new conceptual model of a continuous variation of the characteristics of
quantum-biological processes, coupled with other well-known characteristics 
governing the evolution of self-organized complex structures in nonequilibrium systems, 
guarantee that quantum-mechanically based biological functions
(including human consciousness) do not suffer from the threat of decoherence.

An important corollary from adopting this view is that
among the important drivers for biological evolution (including
the evolution of consciousness), one should not only include 
natural selection of random variations of characteristics,
as Charles Darwin had originally proposed, but also the universal
tendency for self-organized nonequilibrium structures to respond 
directly to their changing and evolving environment.

\section{A NEW VIEW ON THE NATURE OF QUANTUM PROCESSES}

In Zhang (2007), a new ontological view of the quantum processes
is presented.  It is proposed that quantum (as well as classical) 
processes can be understood as a hierarchy of nonequilibrium 
phase transitions, with the lowest hierarchy occurring in a 
``resonant cavity'' formed by the entire matter and energy content 
of the universe. In this formalism, physical laws themselves 
are resonantly-selected, which accounts for the fact that essentially
all quantitative physical laws can be derived from variational
principles, and a correlation exists between the invariance
of transformation properties of physical systems and the conservation
laws these systems obey (Noether's theorem, see, e.g. Neuenschwander
2010).  In this model the values of fundamental constants and properties 
of fundamental particles are determined through a generalized Mach's 
principle (i.e., the numerical values of the fundamental constants
are results of the content and distribution of matter and energy
in the universe).  The existence of a universal preferred reference frame 
is shown to be consistent with the relational nature of physical laws,
such as the relativity theories. The superluminal nature of 
quantum processes in the lowest hierarchy of quantum phase transitions 
is shown to be able to coexist with the universal speed-of-light
limit obeyed by processes in higher hierarchies.  

One important issue 
that has been only briefly mentioned in Zhang (2007), that there is in fact
not a distinctive boundary between a measurement process and a unitary
evolution process, has in the intervening years been much better clarified
by the author. A brief summary on this point is as follows:
In the orthodox interpretation of (nonrelativistic) quantum mechanics, unitary
evolution describes the evolution, according to the Schrodinger equation,
of the probability density representing a quantum system, and the driver for this
evolution is often represented by a constant potential in the Hamiltonian.  
In this picture a quantum measurement is an abrupt interruption of the deterministic 
evolution of the quantum wavefunction, with the result of the measurement being,
in general, uncertain, and with the probability of any particular outcome given
by the overlap of the wavefunction at the start of the measurement
with a new basis function determined by the particular measurement procedure.

However, for the kind of unitary evolution problems which require 
perturbative treatments (see, for example, Sakurai 1985),  including
either the time-dependent perturbation methods (to treat time-dependent potentials), 
or the time-independent perturbation methods (to treat scattering problems),
the solutions of these problems often appear as a continuum of
probability densities $\psi({\bf r})$ (which can be time-dependent as well
in the case of time-varying potential), with the probability density at 
each spatial location having infinitely small absolute value, but accurately
specified relative ratios.  
These perturbatively-treated scenarios, viewed in light of the new 
ontological theory of the quantum processes, are in fact analogous to
the quantum measurement situations, in that an enforced boundary
condition (due to either time-varying potential or the intruding
scattering object) causes the probability distribution of the
incoming wave/mode to either fast-jump (as in scattering) or slow-evolve
(as in time-dependent potential) into a new modal set, and the
results of either these fast or slow evolution processes are in terms 
of the distribution of probability densities, not in terms of the absolute
certainly of the outcome, just as in the traditional quantum
measurement situation (e.g., compare the statement: ``the scattered particle
has x percent of chance to be detected in a given angular range'', as in a 
typical scattering problem, to the statement: ``the quantum particle of 
a given momentum distribution has x percent of chance to be detected 
in a given spatial range'', as in a typical measurement problem).  
In the extreme case, even the non-perturbatively-treated 
unitary evolution situation can be regarded as a continuous
branching-out of the probability densities, thus connecting smoothly
to the scattering and the measurement problems.  Therefore,
the artificial division between the measurement problem and the unitary evolution
problem is only a matter of perspective.

Another essential feature of the new ontological view is that
the quantum mechanical wavefunction now specifies, instead of
a wave of probability or potentiality,  
a realistic matter and energy distribution.  It is only at
the moment of quantum measurement, due to the abrupt nature
of nonequilibrium phase transitions, that this underlying realistic
distribution acquires a new format, with some ``give-and-take'' with
the rest of the universe (Zhang 2007).  These gives-and-takes with the rest of 
the universe underscore the entangled evolution of the entire 
matter and energy content of the universe,
including the evolution of the biological entities from
which consciousness emerges during the natural selection
and evolution process.

\section{NATURAL EMERGENCE OF CONSCIOUSNESS AND FREE WILL DURING
THE COSMIC EVOLUTION OF THE UNIVERSE}

In the new ontological model of the quantum processes we have
summarized in the last section, the zeroth-order hierarchy of
the processes in the quantum universe happen in a substratum
of the continuous distribution of matter and energy content of
the universe (Zhang 2007).  There is an inherent chance or 
indeterministic factor in these processes.  This chance factor 
in the zeroth-order hierarchy is not the same as what 
underlies the ``Heisenberg uncertainty principle'': the latter
in fact is a reflection of the distributed nature of quantum
global modes, and the fact that the characteristics of these modes
cannot be totally ``pinned down'' by simultaneous local measurements (Zhang 2007).
The chance factor we are addressing here is rather related
to the fundamental indeterminism in the phase transition
of an infinite degrees-of-freedom open system (this is similar to
the ``butterfly effect'' in a complex, nonlinear and chaotic system), and if we take
the view that there is substantiality to the wavefunction itself,
including substantiality to the phase of the wavefunction,
then the source of this indeterminism becomes almost
``classical'' (with the understanding that such ``quantum-classicality''
refers to the lowest level in the hierarchy of phase transitions,
and thus is not to be confused with true classical physics which
deals with the higher-hierarchy physical processes).

In true classical physics, the nonequilibrium phase transitions 
happening in open, complex systems have been studied by the late Nobel 
laureate Ilya Prigogine and coworkers as the problem of the formation of 
``dissipative structures'' (Prigogine 1984).  It is well known that for 
an isolated system, the direction of entropy evolution is towards
an increasing degree of macroscopic uniformity.
For open systems at far-from-equilibrium conditions, however,
it often happens that the usual thermodynamic branch of the
solution (i.e. the one that leads to homogeneous distribution)
becomes unstable, and new types of highly organized
spatial-temporal structures emerge spontaneously.  This kind of 
spontaneous structure formation in nonequilibrium systems
has been termed ``nonequilibrium phase transitions'',
and the structures formed ``dissipative structures''
to emphasize the constructive role of dissipation in the
maintenance of these nonequilibrium structures (this constructive
role of dissipation also underlies the well-known
``fluctuation-dissipation'' theorem as applied to
self-organized large-scale fluctuations).
The large-scale coherent orders in open and nonequilibrium systems
are functional as well as architectural.  One of the important functions 
of these ``dissipative structures'' is to greatly accelerate the speed 
of entropy evolution of the parent nonequilibrium systems towards reaching equilibrium
-- even though for certain systems, such as those dominated by self-gravity, an ultimate
thermodynamic equilibrium state does not exist.  The rapidly-produced
entropy by the dissipative structures is promptly exported out
of the system as well, so locally the system could maintain
a highly-organized structure without violating the second
law of thermodynamics (i.e., the fact that the local entropy
production is increased does not contradict with the fact that
the coherent, self-organized structure is maintained since the locally
produced entropy by such a structure is exported out of the system.
So locally the degree of randomness, which we usually associate
with entropy, does not increase, and the organized structure can
thus survive).

In this work, we propose that the emergence of biological
beings -- human beings included, can likewise be viewed as an instance
of the formation of dissipative structures in the complex
universe environment.  In this case the ultimate source of subsequent
nonequilibrium evolution lies in the low-entropy cosmic initial conditions.
One distinguishing feature of the biological beings from other types
of dissipative structures is the the gradual emergence of consciousness.
One naturally wonders where lies the source of this distinguishing
feature.  In what follows, we will present our point of view in
several steps.

First of all, it can be shown that all self-organized dissipative structures 
(including those in the inanimate world) behave ``as if'' they have a ``mind'', 
i.e. these structures organize themselves spontaneously to optimize the achievement of
their predestined mission of accelerating the entropy evolution of the
parent nonequilibrium systems.  In performing this function, these structures
often appear to first ``build a tool'' by forming a complex,
self-sustained morphological pattern which is capable of facilitating entropy 
production.  For self-organized structures the more familiar type of
causality law, i.e., that of local and sequentially-causal, has been
changed to global and mutually-causal (or self-referential).  When a
complex nonlinear system forms a self-organized pattern, it in some
sense has become ``alive''.

An example in this regard is the spiral structure in disk galaxies
such as our own Milky Way and the Whirlpool galaxy M51 (Zhang 1996, 1998; Zhang \& Buta 2010). 
Spiral structures, which are density waves/modes in galaxies,  satisfy many
characteristics shared by other self-organized nonequilibrium dissipative
structures.  For example, a quasi-stationary spiral mode is maintained
by the opposing effect of the spontaneous growth tendency and local dissipation,
with a continuous flux of energy, angular momentum and entropy
through the system carried by the spiral wave itself (Zhang 1998).
It can also be shown that the formation of spiral structures greatly 
accelerates the speed of entropy evolution of a disk galaxy compared 
to that of a uniformly-rotating disk (the so-called ``basic state'', whose
characteristics form the boundary conditions for the emergence
of self-organized spiral modes).  A spiral mode is thus a global instability 
in the underlying basic state of the disk, and the spontaneous emergence 
of the spiral pattern out of an originally featureless, differentially
rotating disk (which is obviously a global symmetry-breaking process) 
happens as long as the disk satisfies certain far-from-equilibrium constraints 
(i.e. the basic state characteristics, including the surface mass density
distribution, the angular momentum distribution, and the random velocity
distribution of the disk-galaxy's star and gas, 
must allow the linear growth rate of a spiral mode to be greater than zero).
The pattern modal characteristics (spiral pitch angle, pattern speed, arm-to-interarm 
ratio) co-evolve with the underlying basic-state characteristics (i.e., the disk
galaxy evolves from the more flattened type, or so-called late Hubble type,
to the more rounded or centrally-concentrated type, or the so-called
early Hubble type, with this evolution itself enabled by the spiral or bar
pattern), since the modal characteristics need to be compatible with the
boundary conditions set by the basic state. 

If we regard biological entities also as ``dissipative structures'',
it is then natural to attribute part of the intelligence 
(or brain-like behavior) of higher biological
beings to the inherent intelligence in the universal meta-laws
governing the evolution of all dissipative structures. 
However, higher biological beings such as humans, besides
possessing the common characteristics of all self-organized
dissipative structures, including features appearing in inanimate systems, 
also possess the additional feature of free will (or intentionality).
We define free will here as the capability
to alter the predestined mechanical evolution of the universe
through deliberate (rather than chance) manipulations by the originators
of the free will.  We argue here that free will is likely to be
the main (or perhaps sole) distinguishing feature that separates
higher biological beings from other types of dissipative structures.
Other features that had previously been assigned competing roles in the definition
of consciousness, such as self-awareness, or else the ability to
behave in a manner that optimize its chance of survival, etc. -- if we look deeper into
what are involved in their functioning, all appear to have a closer
kinship to equivalent functions in lower biological forms and in inanimate
types of dissipative structures.  Take self-awareness, for example.  The
operation of this feature first of all involves the definition of
a self, or else a boundary separating the system from its environment.
It is not hard to see that all self-organized dissipative structures
can successfully be characterized with a boundary between the self
and the environment.  The ``awareness'' part of the characteristics is
equivalent to a system's ability to recognize (either the system or environmental) 
change, and to spontaneously coordinate its behavior to respond to the change, a characteristic 
once again satisfied by all self-organized dissipative
structures.  These systems also have the capability to
transform themselves to best perform their predestined function of optimizing
entropy production, and, in order to achieve this
function, they can even ``build a tool'' to optimize their performance
(i.e. through the self-organization process to build a dissipative structure). 
None of these functionalities require the presence of quantum
interactions or the presence of biological entities.

Free will, or the ability to intentionally and deliberately manipulate the
chance factor in the globally-connected quantum universe, on the other hand, 
does seem to be uniquely associated with biological beings, and with
a biological evolution process that allows this ability to be gradually honed 
to perfection.  Since the ``will part'' of the free will requires a
centralized command center for decision making,
free will in its most developed form is likely to be possessed only 
by higher biological beings with a central processing unit such as a brain,
although a less-developed form of chance manipulation should be present
in lower biological forms as well. 
For free-will to be possible, the governing laws must be genuinely
non-deterministic, in addition to results being non-predictable as in
classical complex systems. The lowest-hierarchy phase transitions in the 
universe resonant cavity clearly satisfy the requirement
of being non-deterministic, because of the singularity nature of the events
originating out of an infinite degrees-of-freedom (or continuous) universe substratum.

The emergence of free will is likely to have 
involved a prolonged evolutionary process: from chance manipulation
of degrees-of-freedom in the continuous substratum by the primordial
forms of biological beings (including single-celled organisms), through
intermediate-complexity biological forms in the tree-of-life (those that
straddle the boundary between operating through conditioned-reflexes, and
deliberate contemplation and decision-making),
and finally arriving at its highly-developed form in humans and other mammals.
It is thus likely that the emergence of human-type consciousness is both
pre-programed in the meta laws of the evolving universe, as an extreme form
of the intelligence inherent in the functionality of dissipative structures,
and also facilitated by the natural selection process.

\section{INSIGHT FROM EASTERN MEDICAL PRACTICES AND PARANORMAL PHENOMENA}

Our current view on the origin and nature of consciousness presupposed
that consciousness is an emergent property of the evolving and interconnected
universe.  It is evolved from a low-level pan-conscious state of lower
biological forms, which learned and perfected the skill to manipulate 
the chance factor in the global phase transitions in the quantum universe.
In this section we look more closely at the functioning of the human
body as revealed by certain eastern medical and spiritual practices, as
well as at certain paranormal phenomena, to seek support for this scenario 
of the gradual evolution of consciousness and its connection with the universe environment.

First we note that the existence of a meridian system 
in the human body has been known for thousands of years in Chinese 
medicinal practice and is the basis of the accupucture and accupressure.
The physical manifestations of the human meridian system
have been confirmed by various modern experiments as channels of low 
resistivity, as well as locations for distinctive sound response upon abrupt impact, 
even though no anatomical features of this system can be visually discerned
-- it is in some sense not unlike the economic forces penetrating a
society even though not everything can be perceived in a concrete form.  
During acupuncture practice, as needle pressure is applied to a specific 
accupucture point on the meridian, this location becomes a temporary 
``command center'', coordinating the responses from other parts of the body,
with signal traveling routes along the meridian system, thus
opening up blocked pathways and invigorating the functioning of the relevant organs.  
Similar flow of energy (or ``chi'') can be induced by 
meditation practices in Hinduism, through focusing on successive ``chakras'' on 
the body and thus enhancing the connectedness of the person as a whole.  
These phenomena suggest that self-organized intelligent behavior can be 
induced in biological entities without the coordinating role of the brain
-- as must have been the case if consciousness is the product of gradual 
evolution from lower biological forms which did not at first
possess a central nervous system.  Consciousness thus originates as
a distributed property of lower biological forms (as is the case for the intelligence
possessed by inanimate dissipative structures), and traces of these
characteristics still manifest in the functioning of the highly-developed
human body even though now the brain and the central nervous system appear
to dominate.

Much of the ``subconscious'' or ``unconscious'' decision-making processes
in human beings involve the substantial participation of the rest of the body.  Much
of the human autonomous nervous system is known to be able to function without
the direct interference of the central nervous system (in fact, much
of the autonomous system's function can be disrupted if the brain exercises
excessive control -- i.e. one's stomach can become upset if one is overly
nervous, that is the time we usually tell the person to relax, and to let
the body own intelligence to take over without the interference of the mind
or the will) .  It is in fact
tempting to speculate that the role of the brain (including its multitudes
of neuron firings) is rather like the role of the government in a
country: it is the executive branch doing the job of running the show,
and yet the source of its intelligence, i.e., the seat of consciousness,
runs much deeper. In our view, the ultimate source of consciousness lies in how
the different components of our body (including all the cells and genes)
inter-relate and interact, and how they are coupled to the universe environment
and to the irreversible evolution of this environment.  Compared to inanimate
self-organized structures such as spiral galaxies or atmospheric
convection cells (the Benard instability), the human body is a highly-developed 
form of spatial-temporal dissipative structure that is constructed
according to quantum-mechanical laws.  The entangled quantum interaction
allows the instantaneous access of human consciousness to resources
that are considered separate and inaccessible by mechanical (or classical)
interactions alone.  This view of consciousness thus leaves open the door
for the explanation of various paranormal and savant behaviors, since
the ability to harness the entangled global resources can conceivably
be naturally gifted to certain fraction of the human population,
or be enhanced though deliberate training.  

\section{IMPLICATIONS ON ARTIFICIAL INTELLIGENCE AND THE FUTURE OF
BIOLOGICAL EVOLUTION}

Another implication of the current proposal is that there can be
no self-aware mechanized artificial intelligence no matter how
advanced the technology becomes, since biological intelligence
is evolved as part of interconnected universal evolution.  
A mechanical computer's CPU will never acquire free will, since it will 
never be able to establish the infinite multitudes of connections 
with the rest of the universe environment (a web of connections
linking all matter and energy in the universe, initiated and evolving
together since the time of the Big Bang), including
entanglement in the phases of the wavefunctions (the
Aharonov-Bohm effect).  However, entities formed through 
semi-biological pathways, such as cloning,
are possible because they tap into nature's established pathways to 
establish these infinite connections.

Recently, the IBM computing machine ``Watson'' has generated a lot of 
press coverage with its stellar performance on the Jeopardy show.  One 
naturally wonders that given another several hundreds years, whether
machines like Watson can be made smart enough to ``think'', to be 
``conscious'' (with an inner awareness), and to possess ``free will''.  
Our answer to these speculations is once again No.  To see why this is so, recall
that consciousness in our view is an emergent property in the
universe environment.  All emergence properties share the characteristics
that it is a kind of ``modal'' property of the underlying system.
You can excite the mode through different types of perturbations,
but the kind of mode that eventually stabilizes for given
boundary condition is highly reproducible (that is why
these modes are called the ``intrinsic modes'' of the underlying
system).  Similar statements can be made for biological entities
under this scenario.  We are not accidents in this universe.  The
characteristics of the universe predetermined the gross characteristics
of our beings even though the details of the events happening around us
are affected both by randomness and by the intervention of our own free will.
A machine like Watson can be invented by IBM with one type
of internal organization, and by others in a European company with a
different type of internal organization.  Both may manifest the
same functionality, but the organizations are up to the choice of
inventors: it is not a product of spontaneous creation of the universe
and not an inevitable element of its continuous evolution.  Another drastic
difference between a computing machine and a biological entity is
that a machine is not fundamentally a ``dissipative structure''
(translates to ``nonequilibrium spatial-temporal mode''), and its functioning
is not regulated by a continuous flux of entropy and energy through
it.  In that sense a computer, no matter how intelligent it appears to
be (the appearance caused by the pre-injected human intelligence), 
and how well it can emulate humans, is fundamentally even less 
intelligent (or less ``alive'') than a spiral galaxy, in that it is 
not dynamically coupled to the environment, and thus cannot naturally
adjust its morphological and dynamical characteristics as the environment
changes (i.e., as the universe evolves).

This view of consciousness as an emergent property of the 
entire biological being also alleviates usual concerns for quantum
decoherence (Tegmark 2000).  Consciousness as a
self-organized, organism-wide phenomenon is in constant exchange and 
in dynamical equilibrium with its environment.  The decoherence
and dissipation tendency is naturally opposed by the self-organization
tendency of naturally-selected resonant modes (i.e., we regard biological
entities and their associated conscious behaviors all as modal
characteristics in the evolving universe resonant cavity). The functioning
of a higher biological entity involves the intricate web of simultaneously
occurring quantum chemical processes, none of these can be viewed as
strictly unitary or strictly measurement processes, but are organized in
a global hierarchical fashion by nature, perfected during the long
history of evolution process.

The recent discovery in the wave-like coordinated behavior in 
photosynthesis processes (see, e.g., Engel et al. 2007, Collini et al. 2010) 
provides another piece of evidence that quantum
self-organized global patterns may be a common underlying feature
in all types of biological processes. Other characteristics of the
biological-entities' functioning, such as magnetic navigation
capability in migrating birds, a growing plant's ability to sense
gravity, light source and seasons, the developing embryo's ability
to decode the DNA information and to assemble complex organs and tissues
and to make these structures function harmoniously in the natural environment,
all point to the connected unfolding of biological entities program
in an intimately participating universe.  The role of consciousness is
not a one-way street, as the brain exercises its free will to shape the
world.  The changing environment also applies its influence on the
continued rewiring of the brain as well as the continued renewal of the
evolving biological species.

The idea of consciousness emerging as a natural process in the
quantum universe also means that consciousness does not perform
the unique role of ``collapsing the wavefunction'' as proposed 
in certain quantum measurement theories. In this regard
consciousness merely joins a multitude of other natural mechanisms 
in participating in global nonequilibrium phase transitions.
The uniqueness of consciousness lies in its highly-evolved ability 
to tap into the non-deterministic potentiality of global phase
transitions so as to make
the carriers of consciousness (i.e. human beings and other higher biological
forms) able to optimally perform their predestined function of
accelerating the entropy evolution of the universe.

\section*{REFERENCES}

Collin, E. et al. 2010, ``Coherently wired light-harvesting in photosynthetic
marine algae at ambient temperature,'' Nature, vol. 463, p. 644

Engel S. et al. 2007, ``Evidence for wavelike energy transfer through
quantum coherence in photosynthetic systems,'' Nature, vol 446, p.782

Neuenschwander, D,E., 2010, ``Emmy Noether's Wonderful Theorem''
(The Johns Hopkins University Press)

Penrose, R., Shimony, A., Cartwright, N., \&
Hawking, S. 1997, ``The large, the small, and the human mind'',
(Cambridge: CUP)

Prigogine, I. 1984, ``Order out of Chaos'' (Shambala)

Sakurai, J.J. 1985, ``Modern quantum mechanics'' (Addison-Wesley)

Stapp, H.P., 2009, ``Mind, matter, and quantum mechanics,'' 3rd edition 
(Berlin: Springer)

Tegmark, M. 2000, ``Importance of quantum decoherence in brain
processes,'' Phys. Rev. E, 61, 4194

Zhang, X. 1996, ``Secular evolution of spiral galaxies. I.  A Collective
Dissipation Process,''  The Astrophy. J., vol. 457, p. 125

Zhang, X. 1998, ``Secular evolution of spiral galaxies. II.  Formation of
quasi-stationary spiral modes,'', The Astrophy. J., vol. 499, p. 93

Zhang, X. 2007, ``On the nature of quantum phenomena'', http://arxiv.org/abs/0712.4297

Zhang, X., \& Buta, R.J. 2010, 
``Density-wave induced morphological transformation of 
galaxies along the Hubble sequence'' (http://arxiv.org/abs/1012.0277)

\end{document}